\documentstyle[12pt,aps,epsf,preprint]{revtex}
\newcommand{\be}{\begin{equation}}
\newcommand{\ee}{\end{equation}}
\newcommand{\ba}{\begin{array}{c}}
\newcommand{\ea}{\end{array}}
\newcommand{\bqa}{\begin{eqnarray}}
\newcommand{\eqa}{\end{eqnarray}}

\begin{document}
\tightenlines
\draft
\title{The Use of Dispersion Relations in the $\pi\pi$ and $K\bar K$
 Coupled Channel System}
\author{Zhiguang Xiao and Hanqing Zheng \footnote{e-mail:
zheng@th.phy.pku.edu.cn}}
 \address{Department of Physics, Peking
University, Beijing 100871,
 \\ People's Republic of China}
\maketitle
 \begin{abstract} Systematic and careful studies are
made on the properties of the IJ=00 $\pi\pi$ and $K\bar K$
coupled--channel system, using newly derived dispersion relations
between the phase shifts and poles and  cuts. The effects of
nearby branch point singularities to the determination of the
$f_0(980)$ resonance are estimated and discussed.
\end{abstract}

\section{Introduction} The physics related to  the  IJ=00 $\pi\pi$
and $K\bar K$ coupled--channel system, especially the property of
the narrow $f_0(980)$ resonance, has been a subject of long and
wide interests both experimentally and theoretically (For an
incomplete list of references, see for example, Ref.~\cite{IZZ} --
\cite{SKN}). In the latest version of Review of Particle
Physics~\cite{RPP00}, the width of the $f_0(980)$ resonance is
estimated to be about 40 to 100 MeV and it is remarked that the
``width determination very model dependent". Previous estimates on
the width of the $f_0(980)$ resonance may essentially be
categorized into two classes: the first is the $S$ matrix or T
matrix pole fit, and the second is from the invariant mass fit in
production experiments. It should be stressed that in a production
experiment the spectrum is highly sensitive to the structure of
the production vertex and therefore the invariant mass fit should
not be considered as very accurate. A numerical calculation to the
formfactor-like quantity in the coupled--channel system
reveals~\cite{lzc} that the behavior of the coupled--channel
form-factor around the $f_0(980)$ region is very fragile: it can
easily generate a peak or a dip, and the difference only comes
from the (bare) production vertices. For the $S$ matrix fit or the
T matrix fit, in the literature various models  have been proposed
to determine the pole position of the $f_0(980)$ narrow resonance.
Different from these approaches we in this paper will use the
dispersion techniques to study the IJ=00 $\pi\pi$ and $K\bar K$
coupled--channel system. The theory of dispersion relation starts
directly from fundamental principles like unitarity and
analyticity and when combining with experimental information the
dispersion theory is predictive. As we will see from the
discussion in this paper, the analytic structure around the
$f_0(980)$ or the $K\bar K$ threshold region is very complicated:
there are branch point singularities which are very close to the
$f_0(980)$ pole, and it is out of question that one has to
carefully examine the influence of these nearby branch point
singularities to the determination of the pole location of the
$f_0(980)$ resonance. Furthermore, the $\pi\pi$ scattering phase
shift $\delta_\pi$, as an analytic function defined in the single
channel unitarity region, is different from the $\delta_\pi$
defined in the coupled--channel unitarity region and the two
quantities have to be treated separately in using dispersion
relations. Though the analytic structure around the $f_0(980)$
region is very complicated, physics related to the $f_0(980)$
resonance can be discussed by using properly established
dispersion relations.

For the simpler case of $\pi\pi$ scatterings in the single channel
approximation, we have in Ref.~\cite{xz} set up a dispersion
relation for the real part (defined in the physical region) of the
T matrix which is related to the experimental observable
$\sin(2\delta_\pi)$ by the following equation,
\begin{equation}\label{ff} \sin(2\delta_\pi)= \rho \left(\sum_i
{\frac{i}{2\rho(z_i) S^{\prime}(z_i)(z-z_i)}} +{\frac{1}{\pi
}}\int_L {\frac{{\rm Im}_L F }{s^{\prime}-z}}ds^{\prime}\right) \
, \end{equation} in which $F\equiv T(1+S)/S$ and $F$ is equal to
$2{\rm Re}T$ in the physical region. In Eq.~(\ref{ff}) the
position of zeros of $S$ on the physical sheet  of the complex $s$
plane are denoted as $z_i$. The discontinuity of $F$ on the
left--hand cut ($l.h.c.$), $L=(-\infty, 0]$ manifests itself in
the left--hand integral on the $r.h.s.$ of the above equation,
where one subtraction to the integral is understood. A very
attractive feature of Eq.~(\ref{ff}) is that it explicitly shows
the contributions  from different types of dynamical
singularities: the resonances  and the left--hand cut (and bound
states or virtual states can be easily included when necessary).
Eq.~(\ref{ff}) has proved to be useful in determining the pole
position of the $\sigma$ resonance~\cite{xz}, especially in
clarifying the role of the left--hand cut in the determination of
the $\sigma$ resonance. That is, the $l.h.c.$ contribution to
$\sin(2\delta_\pi)$ is negative and concave and therefore there
must exist a $\sigma$ resonance to explain the experimental data.

The aim of this paper, as stated above, is to extend the
discussion of Ref.~\cite{xz} to the more realistic case  of the
$IJ=00$ $\pi\pi$ and $K\bar K$ coupled--channel system. We will
set up dispersion representations for physical observables in
which the unitarity cut contributions are dissolved and the
dependence of a physical quantity  on the left--hand cut ($l.h.c$)
becomes explicit.  The method proposed in this paper is rather
general and can in principle be extended to more complicated
situation with more than two channels as well.

The material of this paper is organized as follows: The
Sec.~\ref{c_a} contains the theoretical discussions on the
coupled--channel system. In Sec.~\ref{a_cont} the general property
of the analytic structure of the $\pi\pi$ and $K\bar K$
coupled--channel system is reviewed, including the analytic
continuation and the left--hand cuts. In Sec.~\ref{dr}, dispersion
relations are set up for  analytic functions constructed from the
scattering T matrix. Such analytic functions contain no unitarity
cut. In Sec.~\ref{p_e}, the analytic properties of the newly
constructed functions are analyzed in perturbation theory and in a
simple narrow resonance model. The Sec.~\ref{PD} is devoted to
physical discussions using the dispersion relations founded in
Sec.~\ref{c_a}. In Sec.~\ref{c_d} the physical interpretation of
the dispersion relation, Eq.~(\ref{MAS}), above the $K\bar K$
threshold  is discussed. It is shown that the $\pi\pi$ scattering
phase $\delta_\pi$, defined above the $K\bar K$ threshold contains
a branch point singularity at $s=4m_K^2-4m_\pi^2$. In
Sec.~\ref{couple-single} the dispersion relation for
$\sin(2\delta_\pi)$ in the single channel unitarity region is
established which is a generalization of Eq.~(\ref{ff}) in the
coupled--channel situation. The Sec.~\ref{BW} is for the
application of our method in phenomenology where the property of
the $f_0(980)$ resonance and the influence of the nearby branch
point singularities to the determination of the pole location of
$f_0(980)$ is  carefully examined. In Sec.\ref{general} a general
discussion is given to $f_0(980)$ and its nearby branch point
singularities and Sec.~\ref{PHEN} is devoted to fitting the pole
position of the $f_0(980)$ resonance combining the single channel
and the coupled--channel dispersion relations set up in this
paper.  In Sec.~\ref{subtlety} a subtlety in our procedure is
discussed.  The Sec.~\ref{concl} is for the final conclusion.

\section{The analytic structure of $\pi\pi$ and $K\bar K$
coupled--channel system: General discussions}
\label{c_a}
\subsection{The analytic continuation on the Riemann surface}
\label{a_cont}

As is well known, the  coupled--channel partial--wave  scattering
matrix, $S$, is discontinuous on the unitarity cut. In the two
channel approximation the unitarity cut  includes two kinematical
singularities: the first one starts from $4m_\pi^2$ to $\infty$
whereas the second cut starts from $4m_K^2$ to $\infty$, which
defines a four--sheets Riemann surface. The relation between the
$S$ matrix and the $T$ matrix is,
 \bqa\label{SS} {S_{11}=1+2i\rho_1T_{11}^I}\ , \nonumber\\
 {S_{22}=1+2i\rho_2T_{22}^I}\ ,\nonumber\\
 {S_{12}=2i{\sqrt{ \rho_1\rho_2}T_{12}^I}}\ .
% \nonumber\\ {det S=S_{11}S_{22}-S_{12}S_{21}}\ .
 \cr
 \eqa
 For the $T$ matrix the unitarity relation on the unitarity cut reads,
\bqa\label{uni}
 {\rm Im} T_{11} = T_{11}\rho_1
T^*_{11}\times\theta(s-4m_\pi^2)+T_{12}\rho_2T_{12}^*
\times\theta(s-4m_K^2) \ ,\nonumber \\
{\rm Im} T_{12} =
T_{12}\rho_2 T^*_{22}\times\theta(s-4m_K^2)
+T_{11}\rho_1T_{12}^*\times\theta(s-4m_\pi^2)\ ,\nonumber\\
{\rm Im}
T_{22} = T_{22}\rho_2 T^*_{22}\times\theta(s-4m_K^2)
+T_{12}\rho_1T_{12}^*\times\theta(s-4m_\pi^2)\ ,
\eqa
 where $T_{12}=T_{21}$ has been used.
%and $\rho=diag(\rho_1,\rho_2)$.
The kinematic factors $\rho_1$ and $\rho_2$ are,
$\rho_1=\sqrt{1-4m_\pi^2/s}$ and
$\rho_2=\sqrt{1-4m_K^2/s}$, respectively.
One attempts to extend Eq.~(\ref{uni})  down to the lowest
threshold.  That is, below the second threshold, the above
equations may be written as,
 \bqa\label{sub}
 {\rm Im}T_{11} = T_{11}\rho_1 T^*_{11}\ ,\nonumber\\
 {\rm Im} T_{12} = T_{11}\rho_1T_{12}^*\ ,\nonumber\\
 {\rm Im} T_{22} = T_{21}\rho_1T_{12}^*\ .
 \eqa
 But in general Eq.~(\ref{sub})
would not be correct in the presence of the anomalous threshold
and when the left-hand cut comes across the $4m_\pi^2$ threshold
along the real axis. In the present case of $\pi\pi$ and $K\bar K$ system
there is no anomalous threshold, but the latter does happen. In
the $K\bar K \to K\bar K $ amplitude the left--hand cut
starts from $4m_K^2-4m_\pi^2$ to $-\infty$. As a consequence the third
equation of Eq.~(\ref{sub}) only holds true above $4m_K^2-4m_\pi^2$
along the real axis rather than above $4m_\pi^2$. The $l.h.c.$ in
$T_{11}$ and $T_{12}$ is $(-\infty,0]$, and
$L=(-\infty,4m_K^2-4m_\pi^2]$ for $T_{22}$
(for more information on the location of the $l.h.c.$, see
Ref.~\cite{KS}).

 From Eq.~(\ref{sub}) the analytic continuation of $T$ or
$T^I$ to the second Riemann sheet can be made,~\footnote{We simply
point out that, assuming real analyticity, the space between
$4m_K^2-4m_\pi^2$ and $4m_K^2$ already allows one to make analytic
continuation to the second sheet. One may make use of the K\"allen
and Wightman theorem to refine the analysis, but in here we will
not discuss the subtlety, the content goes beyond the scope of
this paper.}
 \bqa
 T_{11}^{II} &=& {1\over 1+2iT^I_{11}\rho_1 }T_{11}^I\ ,\nonumber\\
 T_{12}^{II} &=& {1\over 1+2iT^I_{11}\rho_1 }T_{12}^I\ ,\nonumber\\
 T_{22}^{II} &=& T_{22}^{I}-2iT_{21}^I\rho_1T_{12}^I
{1\over 1+2iT^I_{11}\rho_1 }\ ,
 \eqa
or in short notations,
\be\label{B2}
 {\bf T}^{II}\equiv {\bf T}^I{\bf B}_{II}={\bf T}^{I}
\left(\matrix{{1 \over 1+2i\rho_1T_{11}^I}& {-2i\rho_1T_{12}^I
\over 1+2i\rho_1T_{11}^I}
 \cr 0 & 1 \cr }\right)\ .
\ee
 From Eq.~(\ref{uni}) one also
obtains the $T$ matrix on the third and sequentially on the fourth sheet,
\be\label{B3}
 {\bf T^{III}}\equiv {\bf T^I B_{III}}={1\over
1+2i{\bf T^I}{\bf}\rho}{\bf T^I}= {\bf T}^{\bf I}
\left(\matrix{{1+2i\rho_2 T_{22}^I\over det S} & {-2i \rho_1
T_{12}^I\over det S} \cr {-2i\rho_2T_{21}^I \over det S} &
{1+2i\rho_1 T_{11}^I \over det S}
 \cr }\right)\ ,\ee
 \be\label{B4}
{\bf T}^{\bf IV}\equiv {\bf T}^{\bf I}{\bf B}_{IV}={\bf T}^{\bf I}
\left(\matrix{1 & 0 \cr {-2i\rho_2T_{21}^I \over
1+2i\rho_2T_{22}^I} & {1 \over 1+2i\rho_2T_{22}^I}
 \cr }\right)\ .
\ee

 Similar discussions on the analytic continuation of the
hadron form-factor,  $F\equiv (F_1,F_2)$, can also be made.
The difference between $F$ and $T$ is that the former is free from
left--hand singularities on the physical sheet by construction.
The
spectral function of $F$ is obtained through the LSZ reduction
formalism,
 \bqa\label{sF}
 {\rm Im} F_{1} = F_{1}\rho_1
T^*_{11}+F_{2}\rho_2T_{12}^*\ ,\nonumber \\ {\rm Im} F_{2} =
F_{2}\rho_2 T^*_{22}+F_{1}\rho_1T_{12}^*\ . \eqa The  analytic
continuation of $F$ is similar to Eqs.~(\ref{B2})--(\ref{B4}),
\be\label{F} F^{II}= F^I{\bf B}_{II}\ ,\,\,\,
 F^{III}=F^I{\bf B}_{III}\ ,\,\,\,
 F^{IV}= F^I{\bf B}_{IV}\ .
 \ee
The continuation to the fourth sheet can either  be obtained by
analytic continuation  from the second sheet or from the third
sheet, which gives of course the same result.
%%%%%%%%%%%%%%%%%%%%%%%%%%%%%%%%%%%%%%%%%%%%%%%%%%
\subsection{The dispersion representation with only left--hand
singularities} \label{dr} To proceed we first notice that the
method used in Ref.~\cite{xz} to derive Eq.~(\ref{ff}), which
requires analyticity of the spectral representation, is not
applicable in the coupled--channel situation. For example, one may
set up the coupled--channel dispersion integral equation for the
form-factor $F$ on the physical sheet (throughout the
 text dispersion integrals are always written in the
un-subtracted form, but possible subtractions are understood),
\be
\label{feq} F={1\over \pi}\int_{4m_\pi^2}^\infty ds'{F(s'){\bf
K(s')}\over s'-s}\ ,
\ee
which defines a 2--dimensional singular
integral equations with Cauchy kernel, where the integral kernel
${\bf K}$, from the above discussion, can be written as,
 \be
{\bf K}(s)= \left(\matrix{\rho_1T_{11}^{II}& \rho_1T_{12}^{II} \cr
0 & 0 \cr }\right)(\theta(s-4m_\pi^2)-\theta(s-4m_K^2))+
\left(\matrix{\rho_1T_{11}^{III}& \rho_1T_{12}^{III}  \cr
\rho_2T_{21}^{III}  & \rho_2T_{22}^{III}   \cr }\right)
\theta(s-4m_K^2)\ .\ee Apparently, ${\bf K}$ is not an analytic
function on the real axis even though $T$ is, hence the method  in
Ref.~\cite{xz} cannot  be used here. Numerical solutions of
Eq.~(\ref{feq}) are searched for in the literature~\cite{lzc}, but
the  numerical solution  is  not unique~\cite{lzc} as a reflection
of the general mathematical theory~\cite{musk}, and the attempt in
picking up the fundamental solution from  others becomes extremely
difficult numerically.

However, as will be shown in the following, the analytic structure of
 $F$ defined in Eq.~(\ref{sF}) can  be studied even though we do not know
how to solve Eq.~(\ref{feq}) analytically. The crucial observation is that,
according to the way of analytic continuation, we have the
following identity,
\be
\label{can} {1\over 2\pi i}\int_C
 \{ {F_I(\tau)\over \tau -z}+ {F_{II}(\tau)\over \tau -z} +
{F_{III}(\tau)\over \tau -z} + {F_{IV}(\tau)\over \tau -z} \}d\tau
\equiv 0\ . \ee where the contour C encircles the right--hand cut
$R\equiv$($4m_K^2-4m_\pi^2$, $\infty$) and along the contour the
complex cut plan is on the left (the contour C would have
encircled the entire unitarity cut if there were no $l.h.c$
intercrossed  the right--hand cut). Since functions $F_I$ --
$F_{IV}$ are analytic on the entire cut plane except for isolated
singularities and except for the left--hand cuts the integral
contour in  Eq.~(\ref{can}) can be deformed, and $F$ can be
expressed as sum of poles and left--hand integrals. The
form-factor may contain bound state poles on the physical sheet
 which correspond
to the bound state poles of $S_I$, and  may contain
 resonances from $B_{II}$ to $B_{IV}$ on the other
sheets (The virtual state, if exist, can also be incorporated into
our formalism easily, but we do not discuss them here for simplicity).
The analytic expression of the scalar form-factor on the physical sheet
can therefore be obtained,
\be
\label{Fz} F(z)=\left[\sum{\tilde F(s_j)\over z-s_j}{\bf
C}(s_j)+\sum_{n=II,III,IV} {F(z_i^{(n)}) \tilde {\bf
B}_{(n)}(z_i^{(n)})\over z-z_i^{(n)}}+{1\over 2\pi i}\int_L
{Disc\left(F{\bf  C}\right)\over z'-z} dz'  \right]/{\bf C}(z)\ ,
\ee
where $L=(-\infty, 4m_K^2-4m_\pi^2]$. In Eq.~(\ref{Fz}) $s_j$ denote the
positions of  bound state poles of $F$ on the physical sheet and
$z_i^{II}$, $z_i^{III}$ and $z_i^{IV}$ denote the position of the
resonance poles  on the second, third and fourth sheet,
respectively. As can be seen from Eqs.~(\ref{B2})-- (\ref{B4}),
they are also zeros of $S_{11}$, det$S$ and $S_{22}$,
respectively. The notations $\tilde F$ and $\tilde {\bf B}$  in
the above equation denote the residues of the corresponding
functions at their pole positions. The denominator ${\bf C}$ in
the above equation is,
 \be\label{C}
 {\bf C}(z) \equiv 1 + {\bf
B}_{II}(z) + {\bf B}_{III}(z) + {\bf B}_{IV}(z)=\left({\matrix{{
2+{1\over S_{11}}+{S_{22}\over {\rm det} S}} &
{-2i\rho_1T_{12}^I\left( {1\over det S}+{1\over S_{11}}\right)}
\cr {-2i\rho_2T_{12}^I\left( {1\over det S}+{1\over
S_{22}}\right)}&{2+{1\over S_{22}}+{S_{11}\over {\rm det }
S}}\cr}}\right)\ .
 \ee
Eq.~(\ref{Fz}) provides an analytic expression of the form-factor
$F$, and  the solution of the integral equation (\ref{feq})
provided that the behavior of the form-factor on the left--hand
cut is known. However, these formulae are highly non-trivial.
Since $F_I$ is free from any left--hand singularity by
construction (though $F$ on other sheets does contain the
left--hand singularities) and contains no resonance pole on the
physical sheet, the residues of $F$ at poles are not free: they
have to be arranged in such a way that on the $r.h.s.$ of
Eq.~(\ref{Fz}), the $l.h.c.$ and the zeros in the numerator (the
term inside the square bracket) and in the denominator, cancel
each other.\footnote{The cancellation has only been verified in
some very simple cases, see Ref.~\cite{xz}.} Though Eq.~(\ref{Fz})
is rather complicated, one simple way to check the correctness of
Eq.~(\ref{Fz}) is to look at the case when the coupled--channel
decouples, that is $T_{21}=T_{12}=0$. In such a case it is easy to
check that the effect of the third sheet is reduced to that of
sheet $II$ and $IV$ and, \be\label{C'} {\bf C}(z) =
2\left(\matrix{{1+S_{11}\over S_{11}}&0 \cr 0 & {1+S_{22}\over
S_{22}} \cr }\right) \ . \ee It is then easy to figure out that
Eq.~(\ref{Fz}) is reduced to two independent single channel
expressions~\cite{xz}.

 In principle, our method presented above can  be
generalized to the case with more than two coupled--channels,
since the key point in deriving the analytic expression,
Eq.~(\ref{can}), can be extended to higher dimensional case. Also,
Eq.~(\ref{Fz}) needs more detailed analysis both theoretically
and phenomenologically. But we will not discuss these topics here.
Instead we will turn to discuss the analytic property of the
coupled channel $T$ matrix.

 The above method can be easily applied to discuss the analytic
structure of the $T$ matrix itself, since the only difference
between $T$ and $F$ is that $T$ is itself discontinuous on the left.
We
have,
 \be
{1\over 2\pi i}\int_C
 \{ {{\bf T}^{\bf I}(\tau)\over \tau -z}+ {{\bf T}^{\bf II}(\tau)\over \tau -z} +
{{\bf T}^{\bf III}(\tau)\over \tau -z} + {{\bf T^{IV}}(\tau)\over
\tau -z} \}d\tau \equiv 0\ .
 \ee
  from which we get
 \be
\label{MAS}{\bf T^I}\left(z\right){\bf C}\left( z\right)+{\bf
\Phi}\left( z\right)-{1\over 2\pi i}\int_L {Disc\left({\bf T^I
C}\right)\over \tau-z} d\tau =0\ ,
 \ee
 where
 \bqa
 \label{TCC} ({\bf TC})_{11}&=&{{1\over
2i\rho_1}\left(S_{11}-{1\over S_{11}}-{S_{22}\over det S}+ {det
S\over S_{22}}\right)}\ ,\nonumber \\
 ({\bf TC})_{22}&=&{{1\over 2i\rho_2}\left(S_{22}-{1\over
S_{22}}-{S_{11}\over det S}+ {det S\over S_{11}}\right)}
\ ,\nonumber\\
 ({\bf TC})_{12}&=&T_{12}\left(1+{1\over S_{11}}+{1\over
 S_{22}}+{1\over det S}\right)\ ,
 \eqa
 and
 \be\label{Phi} {\bf \Phi}(z)=\sum_j{{\rm Res} \left [{\bf TC}(s_j)\right ]\over
s_j-z}+\sum_{i}^{n=II,III,IV} {{\rm Res}\left[{\bf
TC}(z^i_{(n)})\right ]\over z^i_{(n)}-z}\ .
 \ee
 In Eq.~(\ref{Phi})
the first term corresponds to the sum of bound state poles and the
bound state may couple to channel 1 or 2, or both. The residues
can be read off from Eq.~(\ref{TCC}). The second sum on the
$r.h.s.$ of Eq.~(\ref{Phi}) corresponds to the contribution of
resonance poles. It is worth noticing here that not only in
${\bf (TC)}_{22}$ but also in ${\bf (TC)}_{11}$ and
 ${\bf (TC)}_{12}$ the $l.h.c.$ starts from $4m_K^2-4m_\pi^2$.
  One can prove that the matrix function ${\bf TC}$ defined in Eq.~(\ref{TCC})
   satisfies the following
 property,
\be\label{sym} {\bf TC}(\rho_1,\rho_2)={\bf
TC}(-\rho_1,\rho_2)={\bf TC}(\rho_1,-\rho_2)={\bf
TC}(-\rho_1,-\rho_2)\ . \ee

In the case of $\pi\pi$, $K\bar K$ scattering
we assume
there is no bound state pole,\footnote{See however
Refs.~\cite{Janssen95,Locher98,Oller99B}.}
 so  in an explicit form the function
${\bf \Phi}$ defined in Eq.~(\ref{Phi}) can be written as,
 \bqa\label{Phinn}
  \Phi_{11}&=&
-\sum_i{1\over2i\rho_1(z_{II}^i)S'_{11}(z_{II}^i)(z_{II}^i-z)}
-\sum_i {S_{22}(z_{III}^i)/2i\rho_1(z_{III}^i)\over (det
S)'(z_{III}^i)(z_{III}^i-z)}+\sum_i{det
S(z_{IV}^i)/2i\rho_1(z_{IV}^i)\over S'_{22}(z_{IV}^i)(z_{IV}^i-z)}
\, \nonumber
\\
\Phi_{22}&=&\sum_i{det S(z_{II}^i)/2i\rho_2(z_{II}^i)\over
S'_{11}(z_{II}^i)(z_{II}^i-z)} -\sum_i
{S_{11}(z_{III}^i)/2i\rho_2(z_{III}^i)\over(det
S)'(z_{III}^i)(z_{III}^i-z)}-\sum_i{1\over2i\rho_2
(z_{IV}^i)S'_{22}(z_{IV}^i)(z_{IV}^i-z)} \, \nonumber
\\
\Phi_{12}&=&\sum_i {T^I_{12}(z_{II}^i)\over
S'_{11}(z_{II}^i)(z_{II}^i-z)}+\sum_i{T^I_{12}(z_{III}^i)\over
(det S)'(z_{III}^i)(z_{III}^i-z)}+\sum_i{T^I_{12}(z_{IV}^i)\over
S'_{22}(z_{IV}^i)(z_{IV}^i-z)}\ .
 \eqa

\subsection{The analytic property of the function TC}
\label{p_e}
A question naturally arise at this moment is that what
we have done by transforming the matrix ${\bf T}$ to ${\bf TC}$
as expressed by Eq.~(\ref{TCC})?
The answer is of course that under such a transformation the
matrix function ${\bf TC}$ is real analytic on $R$.  The
imaginary part of ${\rm\bf TC}$ in the region
$(4m_\pi^2, 4m_K^2-4m_\pi^2)$ comes solely from the dynamical effects of the
$l.h.c.$, as
${\rm\bf TC}(\rho_1,\rho_2)={\rm\bf TC}(-\rho_1,\rho_2)$ in this region.
To see this
more clearly one may perform a perturbation expansion,
for example a chiral expansion, on ${\bf TC}$. Up to
$O(p^4)$ term, one may write,
 \be\label{p11}
({\bf TC})_{11}={1\over 2i\rho_1}\left(8i\rho_1(T^{(2)}_{11}+T^{(4)}_{11})
+8\rho_1^2T_{11}^{(2)2}+
8\rho_1\rho_2T_{12}^{(2)}T_{21}^{(2)}\right)+O(p^6)\ ,
\ee
 where the superscripts in the small brackets denote the order of
chiral expansion in powers of $p^2$.
 Along the real axis above the second threshold, one can use the perturbation
 unitarity
relation to recast Eq.~(\ref{p11}) as,
 \be
 ({\bf TC})_{11}=4{\rm Re}T_{11}^{(2)+(4)}+O(p^6)\
\ee
 which is indeed real.
Below the second threshold but above the first threshold one uses
the first equation in Eq.~(\ref{sub}) and obtains,
 \be \label{p11s}
 ({\bf TC})_{11}=4{\rm Re}T_{11}^{(2)+(4)}
 -4i\rho_2T_{12}^{(2)}T_{21}^{(2)}+O(p^6)\ ,
 \ee
which is again real analytic.  One should not draw a conclusion
from  Eq.~(\ref{p11s}) that $({\bf TC})_{11}$ is real analytic
down to $4m_\pi^2$. A close look at Eq.~(\ref{TCC}) reveals that
$({\bf TC})_{11}$ contains the  $l.h.c.$ up to $4m_K^2-4m_\pi^2$
which comes from sheet $III$ and sheet $IV$ of the Riemann surface
defined by the $S$ matrix, but these cuts do not show at order
$O(p^4)$. In fact, it is straightforward to check the appearance
of the $l.h.c.$ when we expand $({\bf TC})_{11}$ to  $O(p^8)$,
 \bqa\label{p118}
({\bf TC})_{11}=&&4T_{11}-4i(\rho_1T_{11}^{2}+\rho_2T_{12}T_{21})
-8(\rho_1^2T_{11}^3+\rho_1\rho_2T_{11}T_{12}T_{21}
+\rho_2^2T_{12}T_{22}T_{21})\nonumber \\
&&+8i(2\rho_1^3T_{11}^4+3\rho_1^2\rho_2T_{11}^2T_{12}T_{21}
+2\rho_1\rho_2^2T_{11}T_{12}T_{22}T_{21}+T_{12}T_{21}\rho_2^2
(T_{12}T_{21}\rho_1+2T_{22}^2\rho_2))\nonumber \\ &&+O(p^{10})\ .
\eqa The $l.h.c.$ contribution firstly shows in the third term
inside the second bracket on the $r.h.s.$ of the above equation.
The lowest order contribution it contains is $\sim O(p^4){\rm
Im}T_{22}$. It is easily understood that this term can not be
cancelled by sequential terms in the chiral expansion since the
latter are at least of order of $O(p^6){\rm Im}T_{22}$.

For $({\bf TC})_{22}$ we have
 \be\label{p22}
({\bf TC})_{22}={1\over 2i\rho_2}\left(8i\rho_2T^{(2)+(4)}_{22}
+8\rho_2^2T_{22}^{(2)2}+
8\rho_1\rho_2T_{12}^{(2)}T_{21}^{(2)}\right)+O(p^6)\ ,
\ee
which is equally well written as
\be
 ({\bf TC})_{22}=4{\rm Re}T_{22}^{(2)+(4)}+O(p^6)\
\ee
 above the second threshold, and
\be \label{p22s}
 ({\bf TC})_{22}=4{\rm Re}T_{22}^{(2)+(4)}-4i\rho_2T_{22}^{(2)2}+
 4i({\rm Im}T_{22}^{(4)}-\rho_1T_{12}^{(2)}T_{21}^{(2)})O(p^6)\
 \ee
when $s<4m_K^2$. The difference between here and the 11 channel is
that in here the $l.h.c.$ already appears in the physical sheet
defined by $S$. For completeness we also list result for $({\bf
TC})_{12}$,
 \be\label{p12}
({\bf TC})_{12}=4(T^{(2)}_{12}+T^{(4)}_{12})
-4i(\rho_1T_{11}^{(2)}T_{12}^{(2)}+
\rho_2T_{12}^{(2)}T_{22}^{(2)})+O(p^6)
\ee
 and the $r.h.s.$ is equal to $4{\rm Re}T_{12}^{(2)+(4)}+O(p^6)$
when $s>4m_K^2$ and equals to $4{\rm
Re}T_{12}^{(2)+(4)}-4i\rho_2T_{12}^{(2)}T_{22}^{(2)}+O(p^6)$ when
$4m_K^2-4m_\pi^2<s<4m_K^2$.

From the above discussion it is realized that $({\bf TC})_{11}$ indeed
contains
a left--hand branch point singularity at $s=4m_K^2-4m_\pi^2$.
The perturbative expansion is just used for pedagogical reasons.
Of course the chiral expansion is badly violated in the energy region around
the $K\bar K$ threshold, and the perturbative results should not
be used to argue in favor of  the smallness of the left--hand cut contributions.

It is also helpful to understand more about the property of the
function ${\bf TC}$ by examining how it behaves in some simple
models. For example in a simple Breit--Wigner narrow resonance
model, one may construct the $S$
matrix satisfying unitarity:%~\cite{Zou94B}
\bqa\label{zb}
 T_{11}&=&\frac{e^{2i\phi}-1}{2i\rho_1}+\frac{g_1e^{2i\phi}}
 {M_R^2-s-i(\rho_1g_1+\rho_2g_2)}\ ,\nonumber \\
  T_{12}&=&\frac{\sqrt{g_1g_2}e^{i\phi}}
 {M_R^2-s-i(\rho_1g_1+\rho_2g_2)}\ ,\nonumber \\
   T_{22}&=&\frac{g_2}
 {M_R^2-s-i(\rho_1g_1+\rho_2g_2)}\ ,
 \eqa
where $g_1$, $g_2$  are coupling constants and $M_R$ is the bare
mass parameter.
 In this model the background phase
is simulated by the function $\phi(s)$ with
 \be\label{bgd}
e^{i\phi}=\frac{\alpha(s)+i\rho_1\beta(s)}
{\alpha(s)-i\rho_1\beta(s)}\ ,
\ee
in which $\alpha(s)$ and $\beta(s)$ are real polynomials.
It is easy to verify that the background phase as defined in
Eq.~(\ref{bgd}) obeys the requirement of real analyticity.

The $T$ matrix defined by Eq.~(\ref{zb}) contains both the
$l.h.c.$ (from 0 to $-\infty$) and the right--hand cuts, due to
the presence of the kinematic factors. However a simple algebraic
calculation reveals that the function ${\rm TC}$ is a matrix of
real rational functions, that is, it is analytic on the entire $s$
plane with only isolated singularities. In this sense the
structure of the $S$ matrix defined by Eqs.~(\ref{zb}) and
(\ref{bgd}) is topologically trivial. This observation, which
follows from Eq.~(\ref{sym}) also applies to more general case of
the K matrix parametrization  when K takes the form of rational
functions. Though topologically trivial, simple models as
discussed above can still be helpful in revealing the qualitative
picture of the $\pi\pi$ and $K\bar K$  coupled--channel system.

\section{The analytic structure of $\pi\pi$ and $K\bar K$
coupled--channel system: The two $\delta_\pi$s}
\label{PD}
\subsection{The $\pi\pi$ scattering phase in coupled--channel unitarity region}
\label{c_d}

In the physical region above the second threshold, there is the
well known parametrization of the scattering $S$ matrix,
 \be\label{main}
 S=\left( \matrix{ {\eta {e^{ 2i\delta_{\pi}}}}, &
{i\sqrt{ 1-\eta^2}{e^{i\left(\delta_{\pi}+\delta_K\right)}}}\cr
{i\sqrt{ 1-\eta^2}{e^{i\left(\delta_{\pi}+\delta_K\right)}}}, &
{\eta {e^{ 2i\delta_K}}}\cr}\right)\ .
 \ee
By using it Eq.~(\ref{TCC}) can be recasted as \bqa\label{TCC1}
({\bf
TC})_{11}&=&{1\over\rho_1}{\eta^2+1\over\eta}\sin2\delta_\pi\
,\nonumber \\ ({\bf
TC})_{22}&=&{1\over\rho_2}{\eta^2+1\over\eta}\sin2\delta_K \ , \\
({\bf TC})_{12}&=&{(1-\eta^2)^{1\over 2}\over\sqrt{
(\rho_1\rho_2)}}\left(\cos(\delta_\pi+\delta_K)+{1\over
\eta}\cos(\delta_\pi-\delta_K)\right)\ , \nonumber
\eqa
 and from
Eq.~(\ref{MAS})  we obtain,
 \bqa\label{MAIN}
 {\sin
2\delta_\pi}=-{\eta\over 1+\eta^2}\rho_1\left(\Phi_{11}(s)-{1\over
2\pi i}\int_L {Disc\left({\bf T^I C}\right)_{11}\over z-s}
dz\right)\ ,\cr
{\sin 2\delta_K}=-{\eta\over
1+\eta^2}\rho_2\left(\Phi_{22}(s)-{1\over 2\pi i}\int_L
{Disc\left({\bf T^I C}\right)_{22}\over z-s} dz \right)\ ,\cr
\cos(\delta_\pi+\delta_K)+{1\over
\eta}\cos(\delta_\pi-\delta_K)=-{\sqrt{\rho_1\rho_2}\over
(1-\eta^2)^{1\over2}}\left(\Phi_{12}(s)-{1\over 2\pi i}\int_L
{Disc\left({\bf T^I C}\right)_{12}\over z-s} dz \right) \ .
\eqa
In above equations the function ${\bf \Phi }$  is a  rational
function with real coefficients and the left hand integrals are also
real analytic functions due to the property of  real analyticity
of the $T$ matrix and $L=(-\infty,4m_K^2-4m_\pi^2]$.

 It is convenient to read  from Eq.~(\ref{MAIN})  that
 $\sin(2\delta_\pi)$
as an analytic
function on the entire physical sheet, after analytic continuation,
contains
the $l.h.c.$ starting from $4m_K^2-4m_\pi^2$ to the left of the real axis.
It is easy to understand that
 $\sin(2\delta_\pi)$ and $\eta$ have this  $l.h.c.$ separately in such a way
 that the $l.h.c.$ starting from $4m_K^2-4m_\pi^2$
 in their combination, $S_{11}$,  cancels.
 The appearance of such a $l.h.c.$
 can be clearly seen from the following  expressions for
 $\sin(2\delta_\pi)$ and $\eta$ which are appropriate
 for the analytic continuation,
 \bqa
 \sin(2\delta_\pi)&\equiv&\rho_1{\cal F}_2=
 {1\over 2i}({1\over \eta}S_{11}-{1\over S_{11}}\eta)\ ,
\\ \eta&=&\sqrt{S_{11}S_{22}/ detS}\ , \eqa from which we read
off, \be {\cal F}_2^{III}={\cal F}_2^I\ ,\,\,\,\eta^{III}=\eta^I\
, \ee and \be {\cal F}_2^{II}={\cal F}_2^I\
,\,\,\,\eta^{II}=1/\eta^I\ , \ee which are correct when
Eq.~(\ref{uni}) and Eq.~(\ref{sub}) are correct, respectively.
From these properties it is  realized that on the real axis ${\cal
F}_2$ only contains the $l.h.c.$, and $\eta$ only contains the cut
in the region $[4m_K^2-4m_\pi^2, 4m_K^2]$ (where $\eta$ is a pure
phase) and the $l.h.c.$ on the left. These look good at first
glance for  setting up dispersion relations for these quantities.
But $\eta$ contains many branch point singularities (and so does
$\sin(2\delta_\pi)$)
 on the complex $s$ plane (corresponding  to the positions of resonance poles),
 which prevent the usefulness of these dispersion relations.

In another combination of $\eta$ and $\sin(2\delta_\pi)$ which
appears in the
 first equation of Eq.~(\ref{TCC1}),  as we demonstrated before,
 the right--hand cuts cancel.
Eq.~(\ref{MAIN}) is the appropriate one for
 phenomenological discussions, as
similar to the single channel case~\cite{xz}.
For
example, one may use the experimental data of $\delta_\pi$,
$\delta_K$ and $\eta$ in the region above the $K\bar K$ threshold
to study the properties of the widely concerned
$f_0(980)$ resonance.
 Unfortunately, the
 left--hand integrals in the energy region required by
 coupled--channel analysis are very difficult to estimate
theoretically.~\footnote{The perturbation expansion fails badly
here, and the matrix Pad\'e approximation destroys the single
channel unitarity of the $\pi\pi$
 amplitude by giving it the $l.h.c.$ on the physical sheet~\cite{IZZ,GO99}.}
 This shortcoming
 limits the phenomenological usage of our method. But our approach can still
 be helpful in studying
 the coupled--channel system, as will be discussed in
 Sec.~\ref{BW}.
%%%%%%%%%%%%%%%%%%%%%%%%%%%%%%%%%%%%%%%%%%%%%%%%%%%%%%%%%%%%%%%%%%%%%%%%%%%%%%%%%
\subsection{The $\pi\pi$ scattering phase in single channel unitarity region}
\label{couple-single}
 The analytic structure of the $\pi\pi$ phase shift defined in the
 coupled--channel unitarity region (denoted as $\delta_\pi^{(2)}$ later on)
 is  remarkably different from the analytic structure
of $\delta_\pi$ defined in the single channel unitarity region
(denoted as
$\delta_\pi^{(1)}$ hereafter): \be
\sin(2\delta_\pi^{(1)})\equiv\rho_1{\cal F}= {1\over
2i}(S_{11}-1/S_{11})\ . \ee The function ${\cal F}$ defined in
above equation  only contains the left--hand cut starting from
 $s=0$ together with the ordinary unitarity cut starting from
 $4m_K^2$. It now becomes
 clear that the two $\delta_\pi$ are
not analytic continuation to each other though they match at
$s=4m_K^2$. One can then set up a dispersion relation for ${\cal
F}$, \be\label{single} {\cal F}(s)=\sum_i{Res[{\cal
F}(z_i^{II})]\over s-z_i^{II}} +{1\over
\pi}\int_{-\infty}^0\frac{{\rm Im}_L{\cal F}}{s'-s}ds'
 + {1\over \pi}\int_{4m_K^2}^\infty\frac{{\rm Im}_R{\cal F}}{s'-s}ds'\ ,
 \ee
where the poles are only from the second sheet in the absence of
bound state poles.  Eq.~(\ref{single}) is the extension of the
single channel result~\cite{xz}
 in the coupled--channel case.
Comparing with Eq.~(42) of Ref.~\cite{xz}
only the third term is new and the other two terms are the same.
The first integral on the $r.h.s.$ of
 Eq.~(\ref{single}) has been discussed at length in Ref.~\cite{xz},
 and it is  fortunate that the second integral on the $r.h.s.$
 of Eq.~(\ref{single}) can  be estimated from experiments since
 \be\label{rexp}
 {\rm Im}_R{\cal F}=\frac{1}{2}(1/\eta-\eta)\cos(2\delta_\pi^{(2)})
 \ee
 in the coupled--channel region.
%%%%%%%%%%%%%%%%%%%%%%%%%%%%%%%%%%%%%%%%%%%%%%%%%%%%%%%%%%
\section{Phenomenology of the coupled--channel dispersion
relations} \label{BW}
%%%%%%%%%%%%%%%%%%%%%%%%%%%%%%%%%%%%%%%%%%%%%%%%%%%%%%%%%%
\subsection{The $f_0(980)$ pole and the nearby singularities}
\label{general}

From Eq.~(\ref{single}) we see that when $s$ approaches $4m_K^2$
from the lower side the value of the right integral will change
rapidly since $s=4m_K^2$ is its branch point. It is well known
that the branch point will develop a cusp structure to the real
part of the dispersive integral. Also we see in the above
subsection that $\sin(2\delta_\pi^{(2)})$ contains a left--hand
branch point at $4m_K^2-4m_\pi^2$. Both of the two branch point
singularities are $very$ close to the pole position of the
$f_0(980)$ resonance. Therefore it is necessary  to carefully
investigate the influence of the branch point singularities to the
determination of the pole position. In a typical K matrix fit
dynamical singularities rather than poles are simulated by
background polynomials. However the cusp structure of the branch
point singularity should not be well simulated by a smooth
background polynomial, especially when the background polynomial
is being used to cover a large region of $s$. It is even
reasonable to imagine whether such a cusp phenomenon could be
responsible for the sharp rise of $\delta_{\pi\pi}$ near the
$K\bar K$ threshold.\footnote{It is very impressive to notice that
in Ref.~\cite{AMP}, it was carefully tested up to 40 parameters in
the background phase to check whether the background polynomial
can be responsible for such a sharp rise.} However, a numerical
estimate to the right--hand integral in Eq.~(\ref{single})
indicates that the right--hand cut effect is very weak. In
Fig.~\ref{figrhi} we plot the right--hand integral contribution to
${\cal F}$ from two fits found in the
literature~\cite{AMP,Zou94B}, here the integration is performed
from the $K\bar K$ threshold to roughly about $\sqrt{s}\simeq 1.5$
GeV. As we see from Fig.~\ref{figrhi}
 that the contribution of
the right--hand integral is in the right direction to increase
$\sin(2\delta_\pi^{(1)})$ rather rapidly, but the effect is too
small to have any non-negligible influence to the $f_0(980)$ pole.
This can be clearly seen by comparing Fig~\ref{figrhi} with
Fig~\ref{figsing}. In the latter case $\sin(2\delta_\pi)$ jumps
from -1 to +1 near the $K\bar K$ threshold whereas the
contribution from the right--hand integral is one order of
magnitude smaller. When estimating the right--hand integral we
make use of Eq.~(\ref{rexp}) which assumes implicitly two--channel
unitarity. Though the two--channel unitarity no longer holds true
when the $4\pi$ channel and the $\eta\eta$ channel become
important, we expect the order of magnitude estimate remains
valid.

The left--hand integrals appeared in the coupled--channel
equations, though can not be estimated quantitatively, can also be
proven to behave mildly when the narrow $f_0(980)$ resonance is
located on the second sheet (see Eq.~(\ref{dlhc}) and related
discussions).
%%%%%%%%%%%%%%%%%%%%%%%%

\subsection{A combined fit of the coupled--channel  and the
single channel equations} \label{PHEN}

The single channel dispersion relation, Eq.~(\ref{single})
contains only the second sheet pole whereas the coupled--channel
equations, Eq.~(\ref{MAIN})  contain resonance poles on all
sheets, i.e., sheet II, III and sheet IV. In the single channel
approximation~\cite{xz} we have to introduce 4 parameters for each
resonance pole, two of them are related to the pole position and
the another two are related to the residue, or the resonance
coupling to ${\cal F}$. Now in the coupled--channel case each
resonance pole contains 6 parameters, two of them are for the pole
position, two are for the resonance coupling to ${\cal F}$ (for
second sheet poles only) and/or $\Phi_{11}$ (the second sheet pole
coupling to ${\cal F}$ and $\Phi_{11}$ are the same), and the last
two are for the coupling in $\Phi_{22}$. The pole coupling in
$\Phi_{12}$ is not independent as can be verified from
Eq.~(\ref{Phinn}). It is annoying to have so many parameters
associated with a resonance pole. These parameters are in
principle correlated to each other, but no simple relation with
stringent constraint between these terms are found. Therefore in
here we approximate these parameters as uncorrelated in the fit.
However the property of the $f_0^{II}(980)$ narrow resonance
obtained under such an approximation is expected not distorted
much from the real situation, since the sharp rise of the $\pi\pi$
scattering phase near the $K\bar K$ threshold is so deterministic,
as can be seen from the following discussions.

In the $\pi\pi$ and  $K\bar K$ coupled--channel system, following
the conventional wisdom, we assume two second sheet poles. One
corresponds to the $\sigma $ resonance, the another corresponds to
the $f_0(980)$ resonance, in addition we assume there exists
another 3rd sheet pole which simulate all other pole
contributions.\footnote{A shortcoming of the present method is
that we can not distinguish  the third sheet pole from the fourth
sheet pole in Eq.~(\ref{MAIN}).}  The right--hand integral is
estimated using that depicted in Fig.~\ref{figrhi} which is
however found to have only tiny influence.
 We estimate the left--hand
integral in Eq.~(\ref{single}) in the same way as in
Ref.~\cite{xz}. That is we either use $\chi PT$ to evaluate ${\rm
Im}{\cal F}_L$ but with a truncated integral interval at
$-\Lambda^2$ (here we take $\Lambda$= 600MeV, 700MeV, 800MeV and
1GeV), or estimate ${\rm Im}{\cal F}_L$ from the [1,1] Pad\'e
solution.~\footnote{We are not able to estimate the left hand cut
at quantitative level. In a previous version we had fixed ${\rm
Im}{\cal F}_L$ using the Pad\'e approximant, but then we realized
the problem associated with the Pad\'e approximation~\cite{ang}.
Fortunately the pole positions of the $\sigma$ and the $f_0(980)$
resonances are found not sensitive to the uncertainties related to
our different choices of the left hand cut.}

 In the coupled--channel unitarity region, or in
Eq.~(\ref{MAIN}), however, we do not have a reliable method to
estimate the left--hand integrals. We therefore simulate them by a
group of (totally 3) polynomials. In here we restrict ourselves in
a modest range of energy region, $2m_K<\sqrt{s}<1.2GeV$, therefore
the polynomials may take simple and smooth forms, i.e., constants
or linear polynomials which further introduce a few more
parameters. In the absence of narrow 3rd or 4th sheet pole close
to the branch point singularity at $4m_K^2-4m_\pi^2$ the validity
of this approximation  can be justified. Furthermore, the matching
condition, \be\label{match}
\sin(2\delta_\pi^{(1)})|_{s=4m_K^2}=\sin
(2\delta_\pi^{(2)})|_{s=4m_K^2} \ee is used to reduce one
parameter in the $\pi\pi$ channel.

In the single channel unitarity region we use the combination of
the CERN--Munich data and the low energy data from $K_{e4}$
decay~\cite{Ro77}, and especially the new experimental  data from
the E865 Collaboration~\cite{E865}. We take the $K$ meson mass to
be $m_K=(m_{K^{\pm}}+m_{K_s})/2$ in our fit. In the
coupled--channel unitarity region we use the experimental data on
$\delta_\pi+\delta_K$ and $(1-\eta^2_{00})$/4 from
Refs.~\cite{cohen,MaOz}, and the data on $\delta_\pi$ from
Refs.~\cite{grayer,hyams}.\footnote{Such a choice
of data set merely reflects our prejudice
by excluding the results of Refs.~\cite{Lindenbaum,Frogatt}.}
In order to extract the ``experimental"
value of ${\rm\bf TC}$ in the coupled--channel unitarity region we
have to shift the data  to same points of $s$ by proper
extrapolation.
%Also the experimental value of the scattering
%length, $a_0^0=0.26\pm 0.05$ is also included in the fit.
With such a manipulation to the data, we are able to make the fit,
see Fig.~\ref{figsing} and Figs.~\ref{ccf1}--\ref{ccf3} for a
typical example of the fit where the left--hand integrals in the
coupled--channel equations are  approximated by constants. In the
fit we find that, even for a  very simple parametrization form of
the left--hand integral, there may exist different solutions --
the position of the `third sheet pole' is found to be very
unstable. This even occurs when the 3rd sheet pole is correlated
to the 2nd sheet pole via a Breit--Wigner narrow resonance model
like that in Eq.~(\ref{zb}). In the uncorrelated case the
solutions for the 3rd (or 4th) sheet pole is quite arbitrary: The
mass roughly ranges from $0.7\sim 1.2$GeV and the width ranges
from very small values to roughly about $0.6$GeV. This uncertainty
may be partly due to the unsatisfactory quality of the manipulated
data in the coupled--channel unitarity region (see
Figs.~\ref{ccf1}--\ref{ccf3}), but it is amazing to notice that
such an uncertainty for the 3rd/4th sheet pole does exist in the
literature where very different solutions are also found (see the
compilation of the 2000 edition of the Review of Particle
Physics~\cite{RPP00} for more information). Irrespective of the
large uncertainty that the 3rd/4th sheet pole has,  the $\sigma$
pole position and the $f_0^{II}$ pole position are however found
to be rather  stable against changes of the parametrization form
of  cuts and the uncertainties related to the first left hand cut.
The results are summarized in what follows:
 \bqa
&M_\sigma&=450\sim 510{\rm MeV}\ ,\,\,\, \Gamma_\sigma=480\sim
550{\rm MeV}\ ;\nonumber \\ &M_{f_0^{II}}&=981\sim 992MeV\ ,
\,\,\, \Gamma_{f_0^{II}}=32\sim 40{\rm MeV}\ ;\nonumber \\ &
a_0^0&=0.24\sim 0.27\ . \eqa
% \bqa &M_\sigma&=478\sim 500{\rm MeV}\
%,\,\,\,
%\Gamma_\sigma=480\sim 550{\rm MeV}\ ;\nonumber \\
%&M_{f_0^{II}}&=982\sim 990MeV\ , \,\,\, \Gamma_{f_0^{II}}=35\sim
%37{\rm MeV}\ ;\nonumber \\ & a_0^0&=0.24\sim 0.27\ . \eqa
Very similar to the what happens in the single--channel
situation~\cite{xz}, the location of the $\sigma$ resonance is
sensitive to the choice of the scattering length used in the fit.
If the $\chi PT$ result of $a_0^0=0.220\pm 0.005$~\cite{CGL01} is
used as a constraint in the fit we find instead the following
results (here we take the Pad\'e solution of the first left hand
cut for example):
 \bqa
&M_\sigma&=465\sim 468{\rm MeV}\ ,\,\,\, \Gamma_\sigma=610\sim
650{\rm MeV}\ ;\nonumber \\ &M_{f_0^{II}}&=981\sim 983MeV\ ,
\,\,\, \Gamma_{f_0^{II}}=31\sim 33{\rm MeV}\ ;\nonumber \\ &
a_0^0&\simeq 0.221\ . \eqa

 From  the  results list above we find
that the pole position of the $f_0(980)$ resonance is very stable.
Furthermore the inclusion of the $f^{II}_0(980)$ pole has only
modest influence to the pole position of the $\sigma$ resonance by
comparing with the results from  Ref.~\cite{xz}.\footnote{Previous
results on this point can be found in
Refs.~\protect\cite{DP96,OOP98}.}
 The main reason of
the stability is due to the special data point in the upper right
region in Fig.~\ref{figsing} from the CERN--Munich data:
$\sqrt{s}=0.99$GeV with the value of $\delta_\pi=234.0\pm 12.3$
degrees. The center of mass energy of this data point already
exceeds $2m_{K^\pm}$ though it is located below the $K^0{\bar
K}^0$ threshold. In our fit we set the $K$ meson mass as
$m_K=(m_{K^{\pm}}+m_{K_s})/2$ which is above 0.99GeV,  so we count
this point in the single channel unitarity region (notice that
this data point  is very close to $2m_{K^\pm}$). Also considering
 the experimental
error bar for the beam energy, the location and the value of this
data point  contributes the major uncertainty in the determination
of the location of the $f_0(980)$ resonance (mainly affects the
width). Manifestly our approach is sensitive to the experimental
data near the $K\bar K$ threshold region. On the qualitative
level, however, the picture that there exists a very narrow
$f_0(980)$ pole on the second sheet should not be altered, as is
determined unambiguously by the fact that the value of
$\sin(2\delta_\pi)$ at the matching point (see Eq.~(\ref{match}))
is much closer to $+1$ rather than to $-1$.
%%%%%%%%%%%%%%%%%%%%%%%%%%%%%%%%%
\subsection{The subtlety in our approach: if the $f_0(980)$ narrow
resonance does not locate on the second sheet} \label{subtlety}

As pointed out in the above subsection that the data point closest
to the $K\bar K$ threshold is crucial in determining a narrow
resonance on the second sheet, named as the $f_0^{II}(980)$. One
has to be very cautious under such a situation when the fit is
very sensitive to a single data point. For example, if this point
were removed from the data in fitting Eq.~(\ref{single}), or in
other words if the value of $\sin(2\delta_\pi)$ at the matching
point (see Eq.~(\ref{match})) is closer to $-1$ rather than to
$+1$, the fit program would prefer a solution without the narrow
second sheet pole. Instead, it would give a solution roughly at
$M_{f_0^{II}}\sim 0.86$GeV and $\Gamma_{f_0^{II}}\sim 0.16$GeV and
a very narrow third (or 4th) sheet pole, $M\sim 0.96 - 0.98$GeV
and $\Gamma\sim 5 - 35$MeV. Under such a circumstance it is the
3rd (or 4th) sheet pole combing with the $l.h.c.$ in the first
equation in Eq.~(\ref{MAIN}), rather than the narrow second sheet
pole,  being mainly responsible for the sharp rise of the $\pi\pi$
phase shift $above$ the $K\bar K$ threshold. Though rather
academic, it is worth pointing out that if such a scenario would
happen the narrow pole position could not be reliably estimated
within the present approach, because the third (or 4th) sheet pole
will have a very strong influence to the left--hand integrals in
the coupled--channel dispersion equations. To make the point more
clear let us write down the following formula, \be\label{dlhc}
{1\over 2i}Disc({\bf
TC}_{11})=-4\rho_2^2\Delta_{lhc}\frac{|T_{12}|^2} {S_{22}detS^*} +
h.c.\ , \ee where $\Delta_{lhc}$ is defined as \be {\rm
Im}T_{22}=\rho_1 |T_{12}|^2 +
\Delta_{lhc}\theta(4m_K^2-4m_\pi^2-s)\ ,
\ee
in the single channel
unitarity region. We have no knowledge on how to estimate
$\Delta_{lhc}$ and we assume it behaves normally in the sense that
its order of magnitude and the behavior is similar to, say, the
$l.h.c.$ appeared in the single channel equation. Then it is
apparently shown in the above formula that a very narrow resonance
on the 3rd or 4th sheet near the left--hand branch point at
$s=4m_K^2-4m_\pi^2$ will greatly enhance the discontinuity of
${\bf TC}_{11}$ and also the cusp structure of the dispersion
integral of ${\bf TC}_{11}$ provided that the sign is correct
($i.e.$, to give a sharp rise, not a sharp decrease, to
$\sin(2\delta_\pi^{(2)})$). The 3rd or 4th sheet pole itself would
also contribute in  Eq.~(\ref{MAIN}) (but not in
Eq.~(\ref{single})), a combined contribution of the cut and the
pole may therefore explain the sharp rise of
$\sin(2\delta_\pi^{(2)})$. But of course such a scenario is
unambiguously excluded by experiments.

%%%%%%%%%%%%%%%%%%%%%%%%%%%%%%%%%

\section{Conclusion} \label{concl} In this paper we have studied
the IJ=00 $\pi\pi$ and $K\bar K$ coupled--channel system by using
dispersion relations which are set up after a careful analysis to
the analytic structure of the coupled--channel system. The effects
of various cuts on the determination of pole positions are
discussed and estimated which are found to be mild in general.
Especially the role of the left hand cut starting from
$4m_K^2-4m_\pi^2$ is clarified. Especially, the cut has little
influence to the determination of  second sheet poles.

We therefore confirm, at qualitative level, the conventional
wisdom in the widely used K matrix fit to simulate these dynamical
singularities by mild background polynomials. To some extent, our
analysis also justifies the  dynamical assumption  made in
dynamical approaches (see for example,
Refs.~\cite{OO97,Markushin00}) to partly neglect the $t$ channel
exchange forces by assuming them to behave mildly.

We conclude from the discussion in this paper that there exists a
very narrow second sheet pole with mass about $981\sim 992$MeV and
width about $32\sim 40$MeV which is responsible for the sharp
increase of the $\pi\pi$ scattering phase near (mainly below) the
$K\bar K$ threshold. Meanwhile the mass and the width of the
$\sigma $ resonance is estimated to be $M_\sigma=450\sim 510$MeV,
$\Gamma_\sigma=480\sim550$MeV. However, we pointed out that the
width of the narrow second sheet pole is sensitive to the data
near the $K\bar K$ threshold, and therefore future experiments
with more accurate data are called for. We also find that the
possible third sheet pole which is related to the narrow second
sheet pole in the coupled channel Breit--Wigner resonance model
can not be reliably fixed at present stage within our scheme. The
reason is simply because only the data above the $K\bar K$
threshold can contribute to the determination of the third sheet
pole and also the left hand cut starting from $4m_K^2-4m_\pi^2$
pollutes here.

Our results on the pole locations of the $\sigma$ and $f_0^{II}$
resonances are  in qualitative agreement with the results found in
the literature, though the method used in this paper is rather
different from those commonly used.  In our opinion, the IJ=00
$\pi\pi$ and $K\bar K$ coupled--channel system is so interesting
and important and it is always meaningful to study it from
different angles which can bring us more understanding and insight
in the related physics.  Our method may be generalized to others
channels as well. But the dynamical singularities appeared in the
case of unequal mass scatterings are more complicated. For
example, in the case of $\pi K$ scattering we will run into the
more complicated  problem of dealing with the circular cut.

{\it Acknowledgment}: One of the author, H.Z. would like to thank
Professor Chuan--Rong Wang in Fuzhou University for helpful
discussions. This work is supported in part by China National
Nature Science Foundation under grant  No.~19775005.

\begin{newpage}
%%%%%%%%%%%%%%%%%%%%%%%%%%%%%%%%figure%%%%%%%%%%%%%%%%%%%%%%%%%%%%%%%%%%%%%%%
\begin{figure}[htb]
\begin{center}
\mbox{\epsfysize=90mm \epsffile{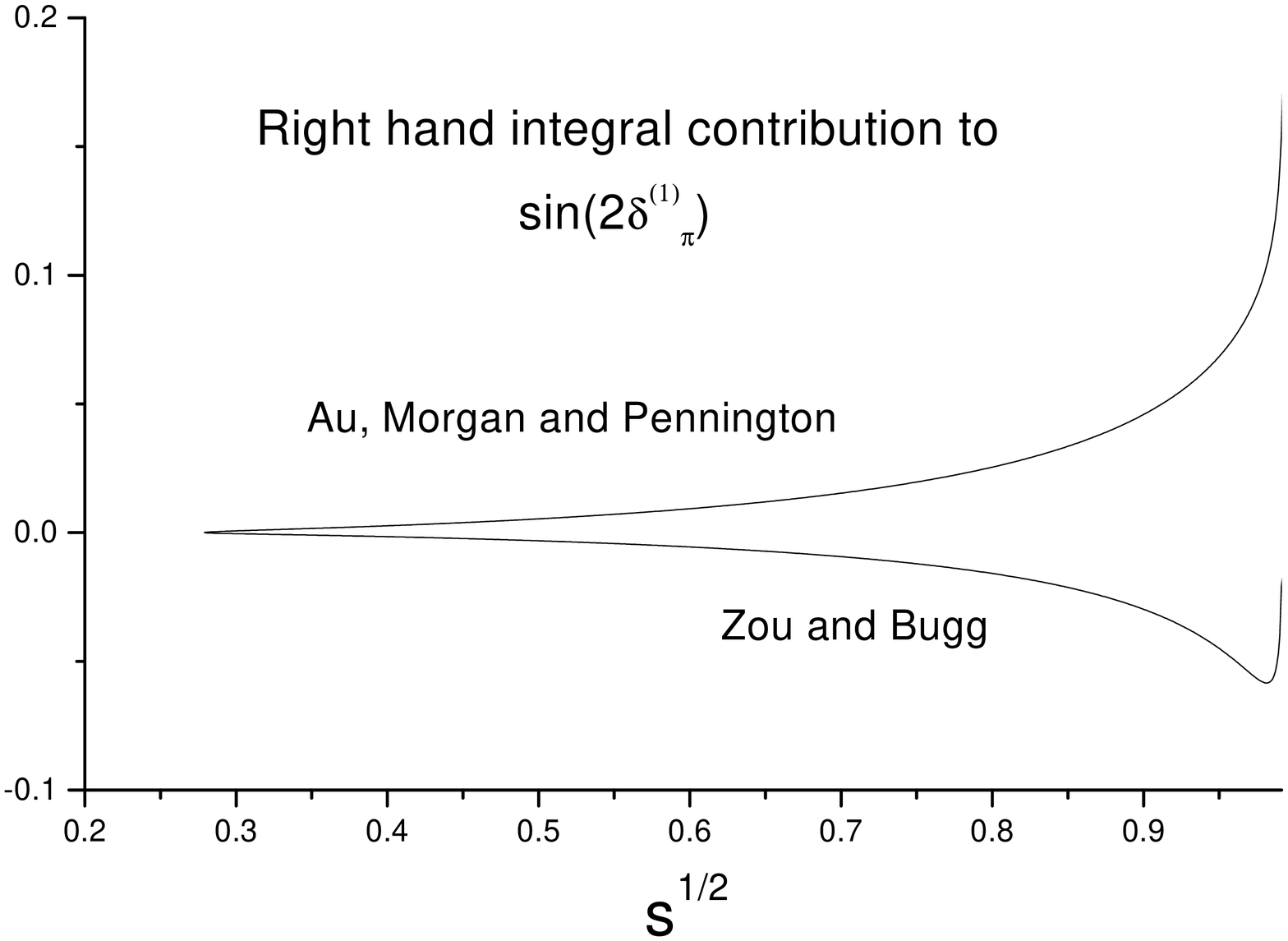}} \hspace*{15mm}
\caption{\label{figrhi}The contribution from the right--hand integral
to \protect${\cal F}$. }
\end{center}
\end{figure}
%%%%%%%%%%%%%%%%%%%%%%%%%%%%%%%%%%%%%%%%%%%%%%%%%%%%%%%%%%%%
%%%%%%%%%%%%%%%%%%%%%%%%%%%%%%%%figure%%%%%%%%%%%%%%%%%%%%%%%%%%%%%%%%%%%%%%%
\begin{figure}[htb]
\begin{center}
\mbox{\epsfysize=90mm \epsffile{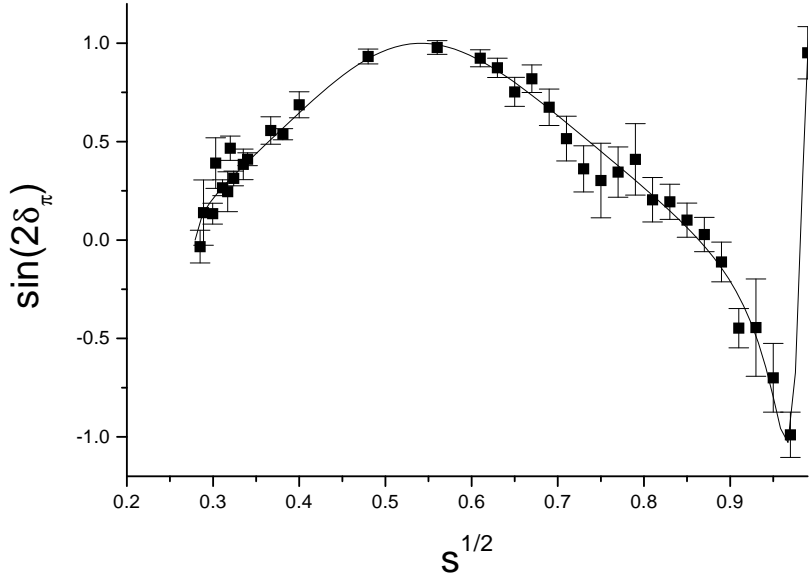}} \hspace*{15mm}
\caption{\label{figsing} A typical fit to $\sin(2\delta_\pi)$
in the single channel unitarity region.}
\end{center}
\end{figure}
%%%%%%%%%%%%%%%%%%%%%%%%%%%%%%%%%%%%%%%%%%%%%%%%%%%%%%%%%%%%
\begin{figure}[hbtp]
\begin{center}
\vspace*{-10mm}
\epsfysize=90mm
\epsfbox{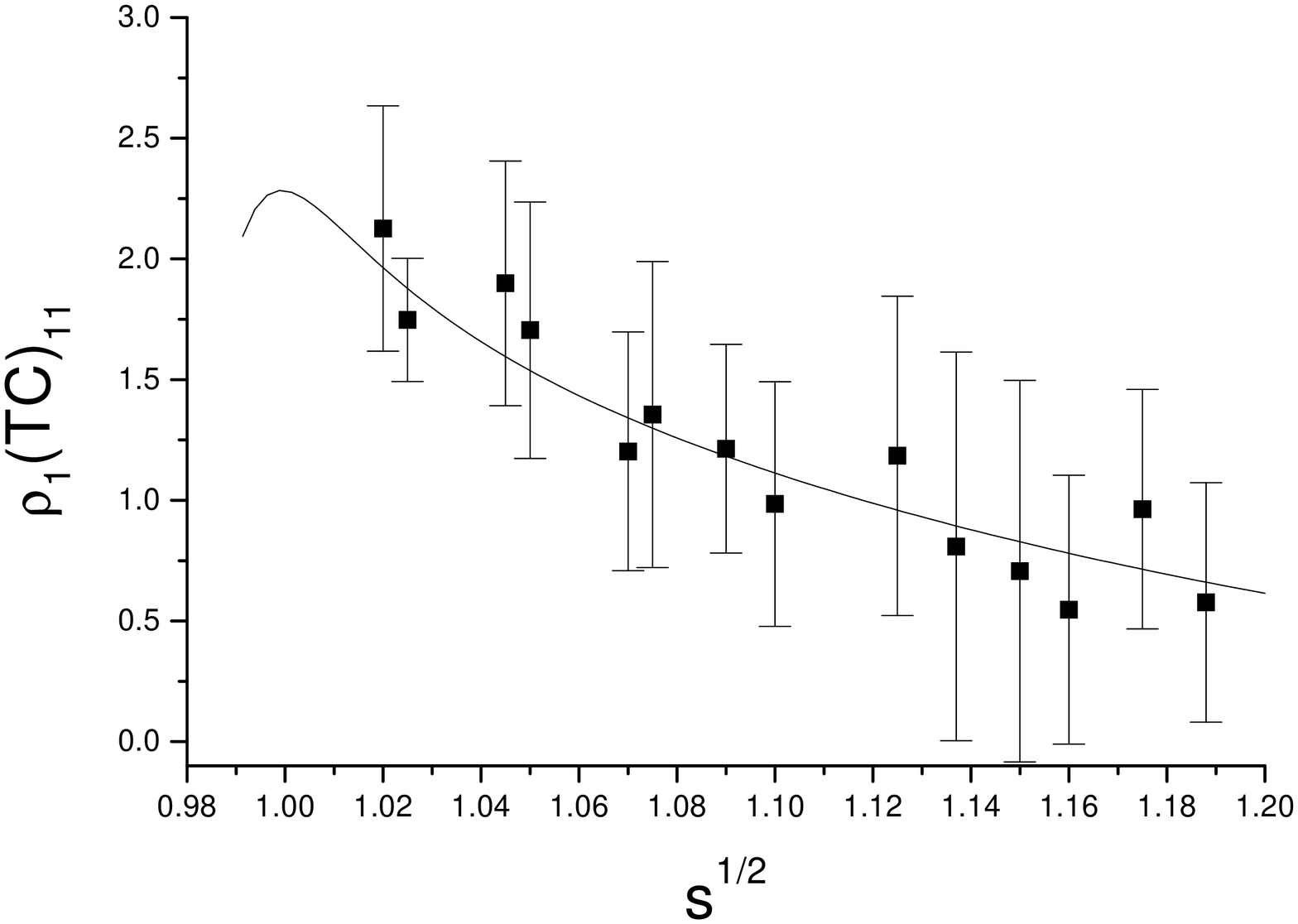}
\vspace*{10mm} \caption{ \label{ccf1}
 A typical fit to ${\rm\bf TC}_{11}$
 in the coupled--channel unitarity region.}
\end{center}
\end{figure}
%%%%%%%%%%%%%%%%%%%%%%%%%%%%%%%%%%%%%%%%%%%%%%%%%%%%%%%%%%%%%%%%%%%%%%%%
\begin{figure}[hbtp]
\begin{center}
\vspace*{-10mm}
\epsfysize=90mm
\epsfbox{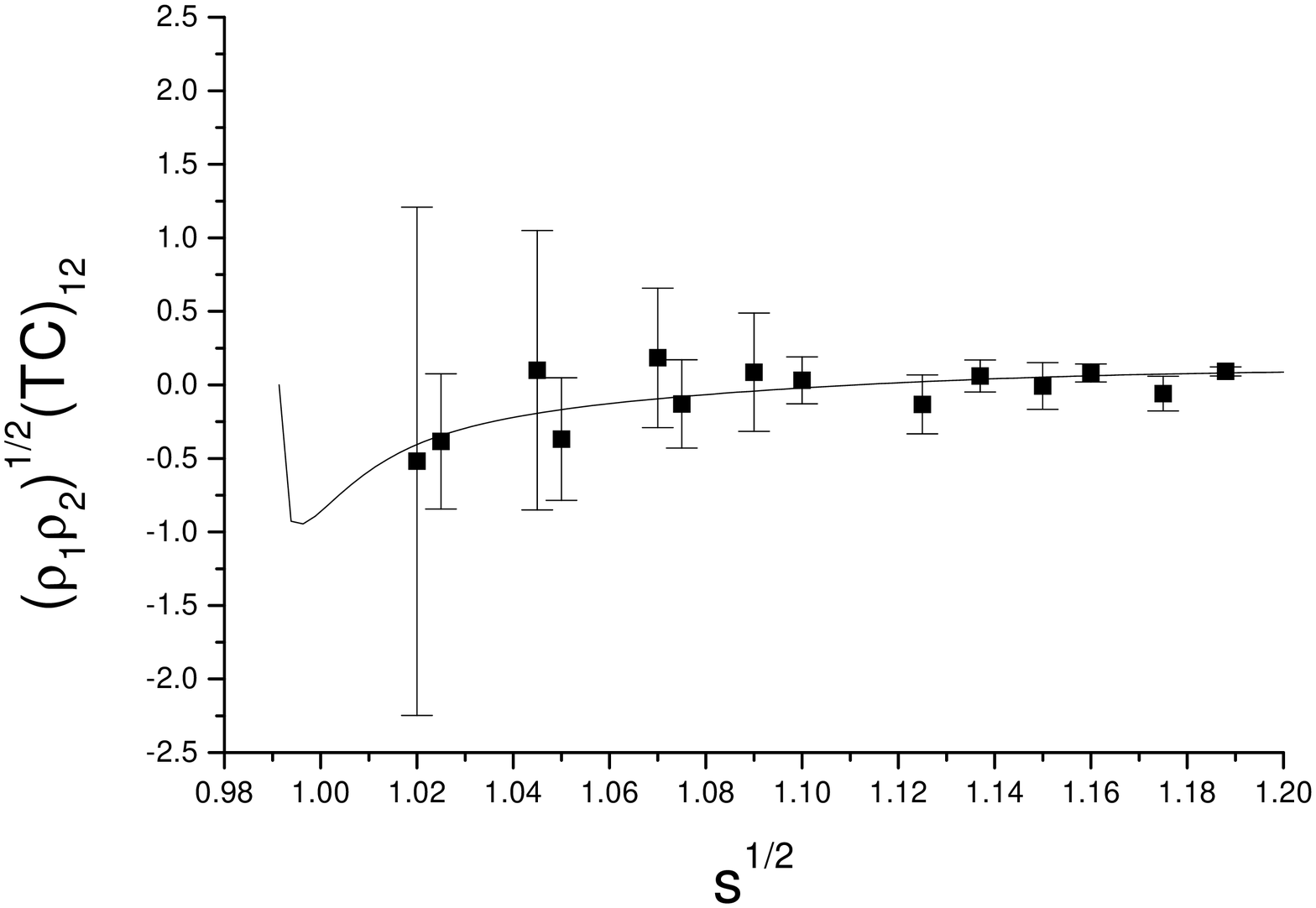}
\vspace*{10mm} \caption{ \label{ccf2}A typical fit to ${\rm\bf TC}_{12}$
 in the coupled--channel unitarity region.
}
\end{center}
\end{figure}
%%%%%%%%%%%%%%%%%%%%%%%%%%%%%%%%%%%%%%%%%%%%%%%%%%%
\begin{figure}[hbtp]
\begin{center}
\vspace*{-10mm}
\epsfysize=90mm
\epsfbox{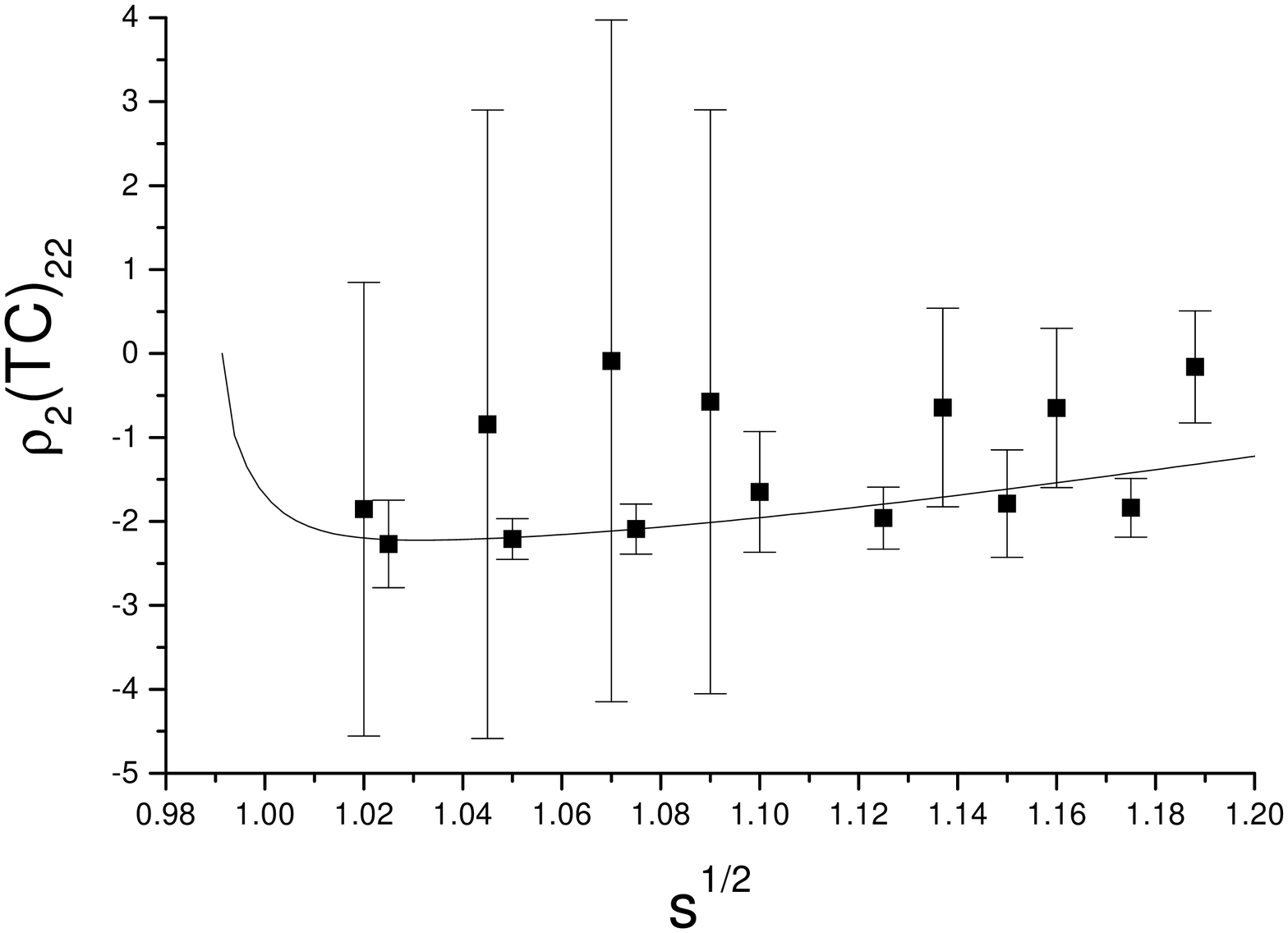}
\vspace*{10mm} \caption{ \label{ccf3}A typical fit to ${\rm\bf TC}_{22}$
 in the coupled--channel unitarity region.
}
\end{center}
\end{figure}
%%%%%%%%%%%%%%%%%%%%%%%%%%%%%%%%%%%%%%%%%%%%%%%%%%%
\end{newpage}

\begin{references}

\bibitem{IZZ}  D.~Iagolnitzer, J.~Zinn-Justin and J.~B.~Zuber,
 Nucl. Phys. {\bf B60}, 233(1973).
\bibitem{AMP}
        K.~L.~Au, D.~Morgan and M.~R.~Pennington,
        Phys. Rev. {\bf D35}, 1633(1987).

\bibitem{Aguilar91}M.~Aguilar-Benitez $et$ $al.$, Z. Phys. {\bf C50}, 405(1991).
\bibitem{Armstrong91}T.~A.~Armstrong $et$ $al.$, Z. Phys. {\bf C51}, 351(1991).
\bibitem{Morgan93}D.~Morgan and
        M.~R.~Pennington, Phys. Rev. {\bf D48}, 1185(1993).
\bibitem{Zou94B}B.~S.~Zou, B.~V.~Bugg, Phys. Rev. {\bf D50}, 591(1994).
\bibitem{Kaminski94}R.~Kaminski $et$ $al.$, Phys. Rev. {\bf D50}, 3145(1994).
\bibitem{Bugg94}D.~V.~Bugg $et$ $al.$, Phys. Rev. {\bf D50}, 4412(1994).
\bibitem{Anisovich94}V.~V.~Anisovich $et$ $al.$,
         Phys. Lett. {\bf B323}, 233(1994).
\bibitem{Amsler94D}C.~Amsler $et$ $al.$, Phys. Lett. {\bf B333}, 277(1994).
\bibitem{Janssen95}G.~Janssen $et$ $al.$, Phys. Rev. {\bf D52}, 2690(1995).
\bibitem{Anisovich95}V.~V.~Anisovich $et$ $al.$,
         Phys. Lett. {\bf B355}, 363(1995).
\bibitem{Amsler95}C.~Amsler $et$ $al.$, Phys. Lett. {\bf B355}, 425(1995);
{\bf B342}, 433(1995).
\bibitem{Alde95B}D.~M.~Alde $et$ $al.$, Z. Phys. {\bf C66}, 375(1995).
\bibitem{Tornqvist96}N.~A.~Tornqvist and M.~Roos, Phys. Rev. Lett. {\bf 76},
1575(1996).
\bibitem{Ishida96}S.~Ishida $et$ $al.$,
                  Prog. Theor. Phys. {\bf 95}, 745(1996).
\bibitem{Bertin97C}A.~Bertin $et$ $al.$, Phys. Lett. {\bf B408}, 476(1997).
\bibitem{Achasov97C}N.~N.~Achasov $et$ $al.$, Phys. Rev. {\bf D56}, 4084(1997).
\bibitem{Locher98}M.~Locher, V.~Markushin and H.~Zheng,
         Euro. Phys. J. {\bf C4}, 317(1998).
\bibitem{Oller99C}J.~A.~Oller and E.~Oset, Phys. Rev. {\bf D60},
074023(1999).
 \bibitem{Oller99B}J.~A.~Oller and E.~Oset, Nucl.
Phys. {\bf A652}, 407(1999).
 \bibitem{Oller99}J.~A.~Oller $et$
$al.$, Phys. Rev. {\bf D60}, 099906(1999).
 \bibitem{Kaminski99}R.~Kaminski, L.~Lesniak and B.~Loiseau, Euro.
Phys. J. {\bf C9}, 141(1999).
\bibitem{Boglione99}M.~Boglione and
M.~R.~Pennington, Euro. Phys. J. {\bf C9}, 11(1999).
 \bibitem{Barberis99D}D.~Barberis $et$ $al.$, Phys. Lett. {\bf
B462}, 462(1999). \bibitem{SKN}Yu.~S.~Surovtsev, D.~Krupa and
M.~Nagy, Acta Phys. Polon. B31 (2000) 2697.
 \bibitem{RPP00} Review of Particle Physics,
                Eur. Phys. J. C15 (2000) 1.
 \bibitem{lzc} W.~Liu, H.~Q.~Zheng and X.~L.~Chen,
  Commun.  Theor. Phys. {\bf 35} (2001) 543.
 \bibitem{xz} Z.~Xiao and H.~Q.~Zheng, hep-ph/0011260, to appear in Nucl. Phys. A.
 \bibitem{KS}
J.~Kennedy and T.~D.~Spearman, Phy. Rev. {\bf 126}, 1596 (1962).
 \bibitem{musk} N.~I.~Muskhelishvili, {\it Singular Integral Equations}, (Moscow,
1946);  J.~K.~ Lu, $Boundary$ $value$ $problems$ $for$ $analytic$
$functions$  World Scientific, Singapore, 1993.
\bibitem{GO99}F.~Guerrero and J.~A.~Oller, Nucl.
         Phys. {\bf B537} (1999) 459.
 \bibitem{ang}Q.~Ang, Z.~G.~Xiao, H.~Zheng and X.~C.~Song,
 hep-ph/0109012, submitted to Commun. Theor. Phys..
 \bibitem{Ro77} L.~Rosselet {\it et
         al.}, Phys. Rev. {\bf D15} (1977) 574.
\bibitem{E865}P.~Truoel $et$ $al.$ (E865 Collaboration),
hep-ex/0012012.
\bibitem{cohen}D.~Cohen, Phys. Rev. {\bf D22},
2595(1980).
\bibitem{MaOz}A.~D.~Martin and E.~N.~Ozmuth, Nucl.
Phys. {\bf B158}, 520(1979).
\bibitem{grayer}G.~Grayer et al.,
Proc. 3rd Philadelphia Conf. on Experimental Meson Spectroscopy,
Philadelphia, 1972 (American Institute of Physics, New York, 1972)
5.
\bibitem{hyams}B.~Hyams et al., Nucl. Phys. B64, 134(1973).
\bibitem{Lindenbaum}S.~J.~Lindenbaum $et$ $al.$,  Phys. Lett. {\bf B274},
492(1994).
\bibitem{Frogatt}C.~D.~Froggatt $et$ $al.$, Nucl. Phys. {\bf B129}, 89(1977).
\bibitem{CGL01}G.~Colangelo, J.~Gasser and H.~Leutwyler, hep-ph/0103088.
\bibitem{DP96}A.~ Dobado and J.~R.~Pelaez, Phys. Rev. {\bf D56} 3057 (1997).
\bibitem{OOP98}J.~A.~Oller,  E.~Oset, J.~R.~Pelaez,
Phys.~Rev.~{\bf D59} 074001 (1999);  Erratum $ibid.$ {\bf D60} 099906 (1999).
\bibitem{OO97}J.~A.~Oller and E.~Oset,  Nucl.~Phys.~{\bf A620} 438 (1997); Erratum
$ibid.$ {\bf A652} 407 (1999).
\bibitem{Markushin00}
V.~E.~Markushin, Eur.~Phys.~J.~{\bf A8} 389 (2000).
\end{references}
\end{document}